\documentclass{iopart}
\usepackage{bm}
\usepackage{graphicx}
\usepackage{leftidx}
\usepackage{pict2e}

\begin{document}

\title[Stability of the SK model near $T=0$]%
      {Stability of the Parisi Solution for the 
        Sherrington-Kirkpatrick model near $T=0$}

\author{A Crisanti$^1$ and C De Dominicis$^2$}

\address{$^1$ Dipartimento di Fisica, Universit\`a di Roma 
              {\em La Sapienza} and  ISC-CNR, 
              P.le Aldo Moro 2, I-00185 Roma, Italy.
        }
\address{$^2$ {\em Institut de Physique Th\'eorique}, CEA -
              Saclay - Orme des Merisiers, 91191 Gif sur Yvette, 
              France
        }
\eads{\mailto{andrea.crisanti@phys.uniroma1.it}, 
      \mailto{cirano.de-dominicis@cea.fr}
     }

\begin{abstract}
To test the stability of the Parisi solution near $T=0$, we study the 
spectrum of the Hessian of the Sherrington-Kirkpatrick model
near $T=0$, whose eigenvalues are the masses of the bare propagators in 
the expansion around the mean-field solution.
In the limit $T\ll 1$ two regions can be identified. 
In the first region, for $x$ close
to $0$, where $x$ is the Parisi replica symmetry breaking scheme parameter,
the spectrum of the Hessian is not trivial and maintains the 
structure of the full replica symmetry breaking state found at higher 
temperatures.
In the second region $T\ll x \leq 1$ as $T\to 0$, the components of the
Hessian become insensitive to changes of the overlaps and the bands 
typical of the full replica symmetry breaking state collapse.
In this region only two eigenvalues are found: a null one and a positive one,
ensuring stability for $T\ll 1$.
In the limit $T\to 0$ the width of the first 
region shrinks to zero and only the positive and null eigenvalues survive.
As byproduct we enlighten the close analogy between the static Parisi replica 
symmetry breaking scheme and the multiple time-scales approach of dynamics,
and compute the static susceptibility showing that it equals 
the static limit of the dynamic susceptibility computed 
via the modified fluctuation dissipation theorem.
\end{abstract}

\submitto{\JPA}

\date{V 2.9.1 2010/12/16 12:23:05 AC}
\maketitle

\section{Introduction:}
The physics of spin glasses is still an active field of research because
the methods and techniques developed to analyze the static and dynamic 
properties have found application in a variety of others fields of the
complex system world, such as
neural networks or combinatorial optimization or glass physics. 
In the study of spin glasses a central role is played by 
the Sherrington-Kirkpatrick (SK) model \cite{SheKir75},
introduced in the middle of 70's, as a mean-field model for spin glasses.
Despite its solution, 
known as the ``Parisi solution'' \cite{Par79,Par80a,Par80b}, 
was found $30$ years ago, some aspect are still far from
being completely understood. 
In this work we discuss the spectrum of the Hessian of the 
fluctuations for the Parisi solution 
in the limit of vanishing temperature, a still not fully explored problem. 

The Hessian spectrum plays a central role non only for the stability of the
Parisi solution of the mean-field SK model, but also for the study of 
finite dimensional systems. Its eigenvalues are indeed the masses of the
``bare'' propagators in the loop expansion about the mean-field limit.
Thus the knowledge of the Hessian spectrum of the SK model is a prerequisite
for any theory obtained from a development about the mean-field limit.

The stability of Parisi solution for the SK model near its critical 
temperature $T_c$, has been established long ago 
\cite{deAlmTho78,DeDomKon83} by 
exhibiting the eigenvalues of the Hessian matrix.
In few words, one has a Replicon band whose 
lowest eigenvalues are zero modes, and a Longitudinal-Anomalous (LA) band, 
sitting 
at $(T_c - T)$, of positive eigenvalues (both with) a band width of order 
$(T_c - T)^2$.
The analysis was partially extended later \cite{DeDomKonTem98} via the 
derivation of Ward-Takahashi identities, showing that the zero Replicon modes 
would remain null in the whole low temperature phase, and hence
would not ruin the stability under loop corrections to the mean-field solution.

Despite these efforts a complete analysis of the stability in the zero 
temperature limit is still missing. Near 
$T_c$ one can take advantage of the vanishing of the order parameter for
$T=T_c$ and expand the free energy, a simplification clearly missing 
close to zero temperature, where the order parameter stays finite.
Moreover the $T=0$ limit is highly non-trivial. All these make the 
derivation of ``effective'' approximations valid for $T\to 0$ a rather
difficult task \cite{CriDeDom10,CriDeDomSar10}.

In this work, anticipating the main results, we show that in the limit 
$T\ll 1$ the spectrum of the Hessian can be divided into two regions.
A first region where the spectrum maintains a structure similar to that found
close to $T_c$, and a second region where only two eigenvalues, 
one null and one positive, are found. In the limit $T\to 0$ the width 
of the first region shrinks to zero, and only the second region survives.

The outline of the paper is as follow. 
In Section \ref{sec:mod-hess} we describe how the Hessian of fluctuations 
associated with the SK model is obtained. 
In Section \ref{sec:simple} we discuss the properties of the Parisi 
solution in the low limit $T\ll 1$ and how these affect the Hessian 
spectrum by considering three simple cases.
In Section \ref{sec:SA} we show how 
spins averages, and response functions, involving any number of spins can be 
computed within the Parisi Replica Symmetry Breaking scheme with a finite 
number $R$ of replica symmetry breaking steps.
In Sections \ref{sec:Repl} and \ref{sec:LA} using the results of
Section \ref{sec:SA} we derive the Hessian spectrum in the $T\to 0$ limit 
for both the Replicon and Longitudinal-Anomalous Sectors.
Finally Section \ref{sec:disc} contains some discussions and conclusions.
The two Appendices contain details on the calculation of spin averages
in the continuous $R\to\infty$ limit, \ref{app:SA-Cont}, 
and the $T\ll 1$ limit, \ref{app:Pankov}.
Finally in \ref{app:FF} for completeness we report
the approach in terms of frozen fields probability distribution functions.

\section{Free energy functional, fluctuations and propagator masses}
\label{sec:mod-hess}
The model is defined by the Hamiltonian \cite{EdwAnd75}
\begin{equation}
  H = -\frac{1}{2}\sum_{i,j} J_{ij}\,s_i\,s_j
\end{equation}
where $s_i=\pm 1$ are $N$ Ising spins located on a regular $d$-dimensional 
lattice and the symmetric bonds $J_{ij}$, which couple nearest-neighbor 
spins only, are random quenched Gaussian
variables of zero mean. 
The variance is properly normalized to ensure a well defined thermodynamic 
limit $N\to\infty$.
To average over the disorder one introduces {\sl replicas}.
After standard manipulations
the free-energy density functional $f$ in the thermodynamic limit is written 
as a function of the symmetric $n\times n$ site dependent 
replica overlap matrix $Q_i^{ab}$ as \cite{BraMoo79}:
\begin{equation}
\label{eq:e1}
e^{-n\,N f/T} = \int\,\prod_{(ab)}\,\prod_i\,dQ_{i}^{ab}\,
              \exp{\cal L}\{Q_i^{ab}\}
\end{equation}
\begin{eqnarray}
\label{eq:e2}
{\cal L}\{Q_i^{ab}\} &=& -\frac{\beta^2}{2}\sum_{\bm p}(p^2+1)
                            \sum_{(ab)}(Q_{\bm p}^{ab})^2
\nonumber\\
&\phantom{=}&
\phantom{=======}
                      +\sum_i\ln{\rm Tr}_{s^a}\,
                   \exp\Bigl(\beta^2\sum_{(ab)}Q_i^{ab}\,s^as^b\Bigr)
\end{eqnarray}
where $Q_{\bm p}^{ab}$ is the spatial Fourier transform of 
$Q_i^{ab}$ with respect to the site index $i$ and $\beta = 1/T$. 
The notation ``$(ab)$'' means that sum is over distinct 
ordered pairs $a < b$ of replicas.

Equations (\ref{eq:e1}) and (\ref{eq:e2}) are the starting point of 
the perturbative expansion around the mean-field theory. One then writes
\begin{equation}
  \label{eq:e3}
  Q_i^{ab}=Q^{ab} + \delta Q_i^{ab}
\end{equation}
where $Q^{ab}$ is the mean-field order parameter, 
and expands ${\cal L}$ in  powers of $\delta Q_i^{ab}$,
\begin{equation}
  \label{eq:e4}
        {\cal L} = {\cal L}^{(0)} + {\cal L}^{(1)} + {\cal L}^{(2)} + \cdots.
\end{equation}
The first term
\begin{equation}
  \label{eq:e5}
        {\cal L}^{(0)} = N\,\left[
                -\frac{\beta^2}{2}\sum_{(ab)}\left(Q^{ab}\right)^2
                      +\ln{\rm Tr}_{s^a}\, 
                        \exp\Bigl({\beta^2\sum_{(ab)}Q^{ab}\,s^as^b}\Bigr)
                        \right]
\end{equation}
gives the free energy density $f$ in the
mean-field limit, and equals that of the SK model.
The second term reads
\begin{equation}
\label{eq:e6}
{\cal L}^{(1)} = -\beta^2\,\sum_i\sum_{(ab)} \delta Q_i^{ab}\,
    \bigl[Q^{ab} - \langle s^as^b\rangle\bigr]
\end{equation}
where
\begin{equation}
\label{eq:e7}
\langle s^as^b\rangle = 
\frac{
  {\rm Tr}_{s^a}\, s^as^b\, \exp\bigl(\beta^2\sum_{(ab)}Q^{ab}\,s^as^b\bigr)
     }
     {
  {\rm Tr}_{s^a}\, 
        \exp\bigl(\beta^2\sum_{(ab)}Q^{ab}\,s^as^b\bigr)
     }.
\end{equation}
The vanishing of ${\cal L}^{(1)}$ yields the stationary condition 
that determines the mean-field value of the order parameter 
$Q^{ab} = \langle s^a s^b\rangle$,
and ensures that tadpoles do not show up in the loop expansion.
Below the critical temperature $T_c$
the phase of the SK model is characterized by a large, yet not
extensive, number of degenerate locally stable states in which the system
freezes. The symmetry under replica exchange is broken
and the overlap matrix $Q^{ab}$ becomes a non-trivial function of replica 
indexes.
In the Parisi parameterization \cite{Par80c} 
the matrix $Q^{ab}$ for $R$ steps of replica exchange symmetry breaking is
divided into successive boxes of decreasing size $p_r$, with $p_0=n$ and 
$p_{R+1}=1$, and elements given
by\footnote{The equality $Q^{ab} = \langle s^as^b\rangle$ that follows from 
  the stationarity condition is valid only for $a\not= b$. 
  For consistency one defines $Q^{aa} = Q_{R+1} = 1$.}
\begin{equation}
 Q^{ab} = Q_{r}, \qquad r = 0,\ldots, R+1
\end{equation}
where $r = a\cap b$ denotes the overlap between the replica $a$ and $b$, and
means that $a$ and $b$ belongs to the same box of size 
$p_r$, but to two distinct boxes of size $p_{r+1}<p_r$. 
The solution of the SK model is obtained by letting $R\to\infty$. 
In this limit the matrix $Q^{ab}$ is described by a continuous non-decreasing
function $Q(x)$ parameterized by a variable $x$,
which in the Parisi scheme is $x\in [0,1]$ and
measures the probability for a pair of replicas to have an overlap not 
larger than $Q(x)$.

The meaning of $x$ depends on the parameterization used for the matrix $Q^{ab}$.
In the dynamical approach \cite{Som81} $x$ labels the relaxation 
time scale $t_x$, so that $Q(x) = \langle s(t_x)\,s(0)\rangle$.
Here the angular brackets denotes time (and disorder) averaging.  
The smaller $x$ the longer $t_x$.
All time scales diverges in the thermodynamic limit but 
$t_{x'}/t_x \to \infty$ if $x > x'$.
To make contact with the static Parisi solution one takes $x\in[0,1]$, 
with $x=0$ corresponding to the largest possible relaxation time
and $x=1^{-}$ to the shortest one. With this assumption one recovers
$Q(0)=0$ and $Q(1^-)= q_c(T)$, the largest overlap. In both cases $Q(1)=1$, since it gives
the self or equal-time overlap. Other choices are
possible, e.g., those used in 
\cite{SomDup84,OppSchShe07,SchOpp08,OppSch08,Schmidt08} to tackle
the $T\to 0$ limit.
We stress however that different choices just give a different parameterization
of the function $Q(x)$, but do not change the physics, since this is given
by the possible values $q$ that the function $Q(x)$ can take and by 
their probability distribution $P(q)$. 
This property is
called {\sl gauge invariance} \cite{Som81,DeDomGabDup82,SomDup84}.
In what follows, unless explicitly 
stated, we take for $x$ the Parisi parameterization.

The quadratic term
\begin{eqnarray}
\label{eq:e8}
{\cal L}^{(2)} &=& -\frac{\beta^2}{2}\sum_p(p^2+1)
                 \sum_{(ab)}\left(\delta Q_{\bm p}^{ab}\right)^2 
\nonumber\\
&\phantom{=}&
\phantom{=======}
               + \frac{\beta^4}{2}\sum_{(ab),(cd)}
               \delta Q_i^{ab}\,
               \delta M^{ab;cd}\,
               \delta Q_i^{cd}.
\end{eqnarray}
defines the ``bare'' propagators of the theory.
This quadratic form in $\delta Q_i^{ab}$ contains the Hessian matrix 
\begin{eqnarray}
\label{eq:e9}
M^{ab;cd} &=& \delta_{(ab);(cd)}^{\rm Kr} - \beta^2\delta M^{ab;cd}
\nonumber\\
          &=& \delta_{(ab);(cd)}^{Kr} - \beta^2\Bigl[
           \langle s^a s^b s^c s^d\rangle - 
           \langle s^a s^b\rangle\langle s^c s^d\rangle 
             \Bigl]
\end{eqnarray}
of the SK model whose eigenvalues rule the stability of 
the mean-field solution, and
give the masses of the ``bare'' propagators.
Terms with 
higher powers of $\delta Q_i^{ab}$ in the expansion (\ref{eq:e4}) 
defines the interaction vertices of the theory.

In the reminder of this paper we shall consider the eigenvalue 
spectrum of the Hessian matrix $M^{ab;cd}$ of the SK model
for the Parisi solution in the very low temperature limit $T\ll 1$.

\subsection{The Hessian $M^{ab;cd}$: 
  the Replicon and the Longitudinal-Anomalous Sectors}

With $4$ replicas the Hessian is characterized by $3$ overlaps.
We can distinguish two different geometries:\\

\noindent (i) The {\sl Longitudinal-Anomalous} (LA) Sector. 
This is characterized 
by the two overlaps $r = a\cap b$ and  $s = c\cap d$ and,
if $r\not= s$, the single cross-overlap 
$t = \max [ a\cap c, a\cap d, b\cap c, b\cap d]$. 
Then we denote the matrix element in the LA Sector as
\begin{equation}
\label{eq:e10}
 M^{ab;cd} = M_t^{r;s}, 
\qquad r,s = 0,1,\ldots R; \ t = 0,1,\ldots R.
\end{equation}
Note that $t=R+1$ if $a=c$ or $a=d$ or $b=c$ or $b=d$. \\

\noindent (ii)
The {\sl Replicon} Sector. In this case 
$a\cap b = c\cap d = r$, and the geometry is characterized by the 
two cross-overlaps
\begin{equation}
\begin{array}{lcl}
 u &=& \max[a\cap c, a\cap d] \\
 v &=& \max[b\cap c, b\cap d]
\end{array} 
\qquad u, v\geq r+1
\end{equation}
For the Replicon Sector the matrix elements are denoted as
\begin{equation}
\label{eq:e11}
 M^{ab;cd} = M_{u;v}^{r;r}, \qquad u,v \geq r+1.
\end{equation}
The element $M_{u;v}^{r;r}$, however, contains contribution from both the 
Replicon and LA Sectors, and one has \cite{DeDomKonTem98b}
\begin{equation}
\label{eq:e12}
M_{u;v}^{r;r} = \leftidx{_{\rm R}}M_{u;v}^{r;r} + M_u^{r;r} + M_v^{r;r} 
                  - M_r^{r;r}
\end{equation}
where the first is the Replicon contribution while the others come 
from the LA Sector.
The latter can be projected out by taking the {\sl double} Replica Fourier 
Transform (RFT) on the cross-overlaps $u,v$:
\begin{equation}
\label{eq:e13}
 M_{\hat{k},\hat{l}}^{r;r} = \sum_{u=k}^{R+1}\sum_{v=l}^{R+1} p_u p_v
         \left[
       M_{u;v}^{r;r} - M_{u-1;v}^{r;r} - M_{u;v-1}^{r;r} + M_{u-1;v-1}^{r;r}
         \right].
\end{equation}
The LA terms indeed cancel in this expression and one can replace
$M_{u;v}^{r;r}$ in the {\sl double} RFT by 
$\leftidx{_{\rm R}}M_{u;v}^{r;r}$.
This in turns implies that the {\sl inverse double}
RFT of $M_{\hat{k},\hat{l}}^{r;r}$ yields the Replicon contribution 
$\leftidx{_{\rm R}}M_{u;v}^{r;r}$  and not $M_{u;v}^{r;r}$.

\section{How things work near $T=0$: simplest cases}
\label{sec:simple}

\begin{figure}
\includegraphics[scale=1.0]{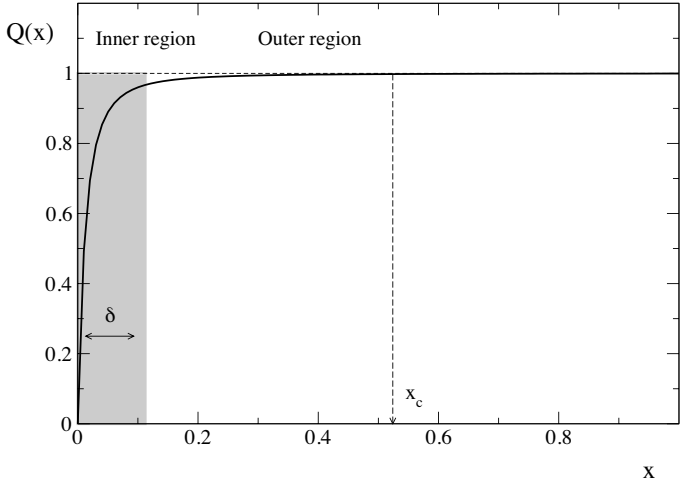}
\caption{Shape of the order parameter function $Q(x)$ for $T\ll 1$
in the Parisi parameterization. The horizontal arrow shows 
the extent of the boundary layer of thickness $\delta\sim T$ as $T\to 0$.}
\label{fig:qx}
\end{figure}

The equation for $Q(x)$ is rather difficult to solve by analytical and/or
numerical methods for $T\to 0$. The origin of this difficulty can be traced 
back 
to the fact that, as the temperature decreases towards $T=0$, the probability
of finding overlaps $Q^{ab}$ sensibly smaller than 
$q_{c}(T) = 1 - \alpha T^2 + {\cal O}(T^3)$, with $\alpha = 1.575\ldots$,
vanishes with $T$ \cite{Som85,CriRiz02}. 
There is however a finite probability $x_c\simeq 0.524\ldots$
that $Q^{ab} \leq q_c(T)$. 
Consequence of this  the order parameter function $Q(x)$ in the Parisi 
parameterization develops for $T\ll 1$ a 
{\sl boundary layer} of thickness $\delta \sim T$ close to $x=0$, 
as shown in Fig.\ref{fig:qx}.
From the Figure we see that for very small $T$ the function $Q(x)$ is slowly 
varying for  $\delta \ll x\leq x_c$. However, in the boundary layer 
$0 < x \leq \delta$,
it undergoes an abrupt and rapid change. 
In the limit $T\to 0$ the thickness $\delta\sim T\to 0$ and the order parameter 
function becomes discontinuous at $x=0$.

Uniform approximate solutions valid for $T\ll 1$ can be constructed 
by using the boundary layer theory, that is by studying the problem 
separately inside ({\sl inner region}) and outside ({\sl outer region}) the
boundary layer \cite{BenOrs99}.
One then introduces the 
notion of the {\sl inner} and {\sl outer} limit of the solution.
The outer limit is obtained by choosing a fixed $x$ outside the boundary layer,
that is in $\delta \ll x \leq 1$, and allowing $T\to 0$. Similarly the
inner limit is obtained by taking $T\to 0$ with $x\leq \delta$. This limit
is conveniently expressed introducing an inner variable $a$, 
such as $a=x/\delta$, in terms of which the solution is slowly varying
inside the boundary layer as $T\to 0$.
The inner and outer solutions are then combined together by matching them 
in the intermediate limit  $x\to 0$, $x/\delta\to\infty$ and $T\to 0$.
The {\sl inner solution} $Q(a)$ is a 
smooth function of $a$ for $T\to 0$ varying between $0$ and $q_c\simeq 1$
\cite{SomDup84,OppShe05,SchOpp08,OppSch08}, similar to $Q(x)$ at finite 
temperature. In the rest of this paper we shall concentrate on
{\sl outer solution} since as $T\to 0$ it covers the overwhelming part 
of the interval $[0,1]$.

The behavior of $Q(x)$ for $T\ll 1$ has strong consequences on
other relevant quantities, such as, e.g, the four-spin correlation entering 
into the Hessian matrix. We shall make this more quantitative in the next 
Sections. 
Here the only feature we whish to retain is that 
in the {\sl outer region} for $T\ll x < x_c$ and $T\to 0$, 
the function  $Q(x)$ is driven closer and closer to $Q(x_c) = q_c(T)$ as 
$T$ approaches zero. 
It can be shown \cite{Pan06}, see also \ref{app:Pankov}, that
for $T \ll x\leq x_c$ and $T\to 0$
\begin{equation}
\label{eq:q-pank}
  Q(x) = 1 - c\, (\beta x)^{-2} + \left(\frac{c}{x_c^2} - \alpha\right)\, T^2
         + {\cal O}\bigl((\beta x)^{-3}, T^3\bigr),
\end{equation}
where $c = 0.4108\ldots$ and $\alpha = \lim_{T\to 0}(1 - q(x_c))/T^2$.
We note that the breakpoint 
$x_c$ depends on $T$.
The dependence is however very weak for low temperatures \cite{CriRiz02}
and the approximation 
$x_c(T) \simeq x_c = 0.524\ldots$ is rather good for $T\sim 0$.
From this expression we see that the variation of $Q(x)$ in the 
{\sl outer region} is
\begin{equation}
  \frac{Q(x_c) - Q(x)}{Q(x)} \simeq c\left(\frac{T}{x}\right)^2
                                \left[1 - \left(\frac{x}{x_c}\right)^2\right]
\end{equation}
so that one can safely take the approximation $Q(x)\sim Q(x_c) = q_c(T)$ as 
$T\to 0$, the error being ${\cal O}(T^2)$ at least.
Going back to $R$ steps of Replica Symmetry Breaking 
this approximation translates into
\begin{equation}
\label{eq:qc}
  Q_r \sim Q_R=q_c(T) = 1 - \alpha T^2 + {\cal O}(T^3),
\qquad T\to 0
\end{equation}
for all $r$ in the {\sl outer region}, that is such that 
$T\ll x(Q_r) = p_r$ and $T\to 0$, or, equivalently, for fixed $r\not= 0$ 
and $T\to 0$.
We shall make this {\sl insensitivity} with respect to the overlaps $r$ 
in the $T\ll 1$ limit more precise in the next Sections. Here we just discuss
the consequence of the {\sl insensitivity} on the elements of the Hessian  
by considering some simple cases.

Suppose the two pairs of replicas are equal:
$(a,b)=(c,d)$. In this case from eq. (\ref{eq:e9}) one 
constructs the simplest Hessian component:
\begin{eqnarray}
\label{eq:e14}
M^{ab;ab} &=& 1 - \beta^2\Bigl[
   \langle(s^as^b)^2\rangle - \langle s^as^b\rangle \langle s^as^b\rangle
                             \Bigr]
\nonumber\\
 &=& 1 - \beta^2\Bigl[1 - \left(Q^{ab}\right)^2\Bigr]
\end{eqnarray}
that for the overlap $a\cap b= r$ gives
\begin{equation}
\label{eq:e16}
M_{R+1;R+1}^{r;r} =1 - \beta^2(1-Q_r^2).
\end{equation}
{\sl Insensitivity} implies that for fixed $r$ and $T\to 0$ we have
\begin{equation}
\label{eq:e17}
M_{R+1;R+1}^{r;r} \sim M_{R+1;R+1}^{R;R} 
                  = 1 - 2\alpha + {\cal O}(T^2),
\qquad T\to 0.
\end{equation}
\begin{figure}
\includegraphics{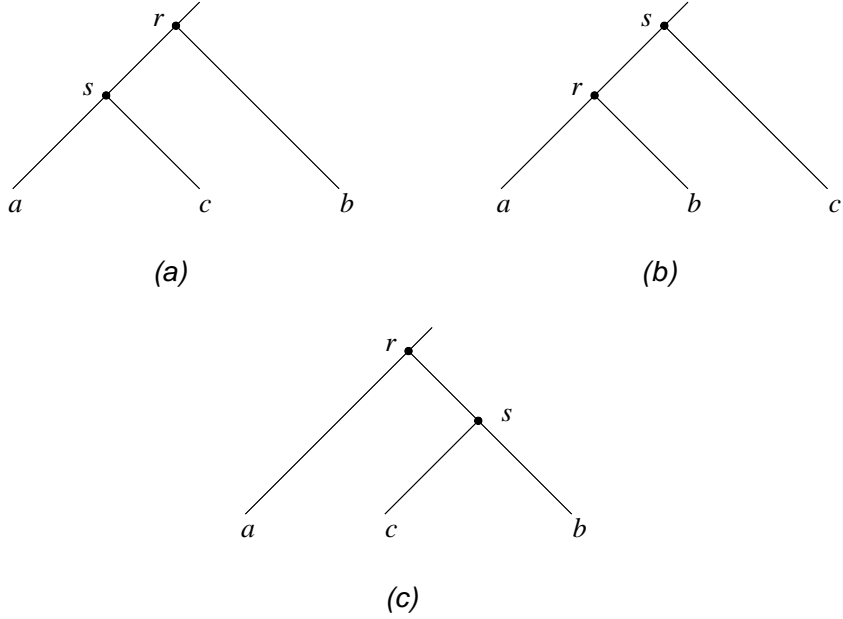}
\caption{Tree configuration for replicas $a,b,c$ with
 $a\cap b= r$.}
\label{fig:three-spin}
\end{figure}
The next simple case is when only three replicas are different, in which case
we have
\begin{equation}
\label{eq:e18}
M^{ab;ac} = -\beta^2\Bigl[
   \langle s^bs^c\rangle - \langle s^as^b\rangle\langle s^as^c\rangle
                                     \Bigr], \qquad b\not= c.
\end{equation}
Ultrametricity imposes that the three replicas $a,b,c$ with $a\cap b=r$ 
can be only disposed as shown in Fig. \ref{fig:three-spin}.
The LA geometries $(a)$ and $(b)$ lead for $T\to 0$ and fixed $r$ and $s$ to
\begin{equation}
\label{eq:e19}
M_{R+1}^{r;s} = -\frac{Q_r - Q_r Q_s}{T^2} \simeq 
                -\frac{Q_R(1-Q_R)}{T^2} = -\alpha + {\cal O}(T^2),
\end{equation}
while the Replicon geometry $(c)$ yields
\begin{equation}
\label{eq:e20}
M_{R+1;s}^{r;r} =
-\frac{Q_s - Q_r^2}{T^2} \simeq 
                -\frac{Q_R(1-Q_R)}{T^2} = -\alpha + {\cal O}(T^2).
\end{equation}
We shall see below that {\sl insensitivity} implies that 
$M_s^{r;r}-M_r^{r;r}\sim 0$, and that all Replicon components vanish.
Then from eq. (\ref{eq:e12}) and (\ref{eq:e20})
it follows
\begin{equation}
\label{eq:e22}
 M_{R+1}^{R;R} = -\alpha + {\cal O}(T^2), 
\qquad T\to 0.
\end{equation}
Similarly from (\ref{eq:e12}) and (\ref{eq:e17}) one obtains
\begin{equation}
\label{eq:e23}
 2M_{R+1}^{R;R}- M_R^{R;R} = 1 - 2 \alpha + {\cal O}(T^2)
\end{equation}
which combined with (\ref{eq:e22}) gives
\begin{equation}
\label{eq:e24}
 M_R^{R;R} = -1 + {\cal O}(T^2),
\qquad T\to 0.
\end{equation}

The general case with four different replicas cannot be reduced to simple forms
and the expression of the four-spin averages is required. This will be derived 
the next Section.

\section{Spin Averages}
\label{sec:SA}
The evaluation of the Hessian components requires the
computation of the four-spin averages $\langle s^a s^b s^c s^d\rangle$
for a generic geometry of the four replicas.
This can be done by introducing the generating function 
\begin{equation}
\label{eq:sa-gf}
{\cal Z}({\bm b}) = \exp \bigl\{nG_{-1}({\bm b})\bigr\}
                  = {\rm Tr}_{s^a}\, 
                        \exp\left({\frac{1}{2}\sum_{ab}\Lambda_{ab}\,s^as^b 
                            + \sum_{a} b_a\,s^a}\right)
\end{equation}
where $\Lambda_{ab}$, equal to $\beta^2 Q_{ab}$ with $\beta = 1/T$ for the 
SK model, is a generic $n\times n$ symmetric matrix with Parisi's block
structure,
\begin{equation}
  \left. \Lambda_{ab} \right|_{a\cap b = r} = \lambda_r, \qquad 
   r = 0,\ldots, R+1.
\end{equation}
Spin averages follow from differentiation
\begin{equation}
\label{eq:sa-spa}
\langle s^a s^b\cdots\rangle = \lim_{n\to 0} 
                  \frac{1}{{\cal Z}({\bm b})}\
\Bigl[
                  \frac{\partial}{\partial b_a}
            \left.\frac{\partial}{\partial b_b} \cdots\Bigr]\ {\cal Z}({\bm b})
            \right|_{b_1=\cdots=b_n = 0}.
\end{equation}
Introducing the ``block indexes'' $a_k$,
\begin{equation}
\label{eq:sa-blki}
  a = (a_0,a_1,\ldots, a_R), \qquad
  a_k = 0,\ldots \frac{p_k}{p_{k+1}} - 1
\end{equation}
where $p_k$, with $n=p_0 > p_1 > \cdots > p_R > p_{R+1} = 1$, 
are the block sizes, the generating function can be written 
as multiple integrals over independent Gaussian variables:%
\footnote{We use Greek letters for summed replica indexes}
\begin{equation}
\label{eq:sa-gfr}
{\cal Z}({\bm b}) = \int {\cal D}_R(\alpha) \prod_{0}^{R} 
                    \exp G_R(b^R_\alpha + b_\alpha)
\end{equation}
where ${\cal D}_{R}(\alpha)$ is the short-hand notation for:
\begin{eqnarray}
  \int {\cal D}_R(\alpha) &\equiv& 
   \prod_{t=0}^R \int \prod_{0}^{t-1} Dz^t_\alpha 
\nonumber\\
&=&
   \prod_{t=0}^R \int \prod_{\alpha_0,\ldots, \alpha_{t-1}} 
              Dz_{\alpha_0,\ldots,\alpha_{t-1}}
\end{eqnarray}
and $z^t_{\alpha} = z_{\alpha_0,\ldots,\alpha_{t-1}}$ 
are independent Gaussian random 
variables of zero mean and variance one:
\begin{equation}
 Dz \equiv \frac{d z}{\sqrt{2\pi}} e^{-z^2/2}.
\end{equation}
The function $G_R(b)$ is the ``free energy'' of a single spin in a field $b$:
\begin{equation}
\label{eq:sa-GR}
  \exp G_R(b) = {\rm Tr}_{s}\, \exp (b\, s) 
                = 2 \cosh b.
\end{equation}
and the frozen  (random) field $b^R_{\alpha}$, given by
\begin{equation}
\label{eq:sa-bR}
 b^R_{\alpha} = \sum_{t=0}^{R} \sqrt{\Delta\,\lambda_t}\, z^t_{\alpha}
\end{equation}
where $\Delta\lambda_t = \lambda_t - \lambda_{t-1}$,
keeps track of the contributions from the various blocks. 

Inserting the form (\ref{eq:sa-gfr}) of ${\cal Z}({\bm b})$ into eq. 
(\ref{eq:sa-spa}), and noticing that differentiation with respect to $b_a$
can be replaced by differentiation with respect to $b^R_a$, we obtain
\begin{eqnarray}
\label{eq:sa-spaR}
\langle s^a s^b\cdots\rangle &=& \lim_{n\to 0} 
                     \frac{1}{{\cal Z}(0)}
\left[
                  \frac{\partial}{\partial b^R_a}
                  \frac{\partial}{\partial b^R_b} \cdots 
\right]
        \int {\cal D}_R(\alpha) \prod_{0}^{R} 
                    \exp G_R(b^R_\alpha)
\nonumber\\
     &=& \lim_{n\to 0} 
                     \frac{1}{{\cal Z}(0)}
        \int {\cal D}_R(\alpha) \prod_{0}^{R} 
                    \exp G_R(b^R_\alpha) 
\nonumber\\
&\phantom{=}& \phantom{======}
\times
\left[\frac{\partial}{\partial b^R_a}\,G_R(b^R_a)\right]
\left[\frac{\partial}{\partial b^R_b}\,G_R(b^R_b)\right]
                      \cdots
\end{eqnarray}

For any given geometry of the replicas $a,b,c\ldots$ the integrals can now 
be performed recursively from scale 
$R$ up to scale $0$. To illustrate the procedure let us consider
\begin{equation}
\label{eq:sa-z0}
{\cal Z}(0) = \int {\cal D}_R(\alpha) \prod_{0}^{R} 
                    \exp G_R(b^R_\alpha).
\end{equation}
The field $b^R_\alpha$ can be written as
\begin{eqnarray}
 b^R_\alpha &=& \sqrt{\Delta\lambda_R}\, z^R_\alpha + 
               \sum_{t=0}^{R-1} \sqrt{\Delta\,\lambda_t}\, z^t_\alpha
\nonumber\\
             &=& \sqrt{\Delta\lambda_R}\, z^R_\alpha + b^{R-1}_\alpha.
\end{eqnarray}
Then splitting out the $z^R_\alpha$-integrals, and 
recalling that $z^R_\alpha$ depends only upon indexes 
$\alpha_0,\ldots,\alpha_{R-1}$, one has:
\begin{equation}
\fl
\label{eq:sa-z01}
{\cal Z}(0) = \int {\cal D}_{R-1}(\alpha) \prod_{0}^{R-1}
         \int Dz^R_\alpha
  \exp\left\{p_R\,G_R(\sqrt{\Delta\lambda_R}\, z^R_\alpha
                           + b^{R-1}_\alpha
                     )
      \right\}.
\end{equation}
This structure suggests introducing quantities $G_r(b)$ as
\begin{equation}
\label{eq:sa-gr}
  \exp \left\{p_r\, G_{r-1}(b^{r-1}_\alpha)\right\} = 
         \int Dz^r_\alpha\,
  \exp\left\{p_r\,G_r(\sqrt{\Delta\lambda_r}\, z^r_\alpha 
                           + b^{r-1}_\alpha
                     )
      \right\}
\end{equation}
so that eq. (\ref{eq:sa-z0}) can be written as
\begin{equation}
\label{eq:sa-z01a}
{\cal Z}(0) = \int {\cal D}_{R-1}(\alpha) \prod_{0}^{R-1} 
           \exp \left\{p_R\,G_{R-1}(b^{R-1}_\alpha)\right\},
\end{equation}
that has the same form of (\ref{eq:sa-z0}) provided $R\to R-1$. 
The entire process can be iterated up to level $0$ and leads to
\begin{equation}
\label{eq:sa-z00}
{\cal Z}(0) = \exp\left\{ p_0 G_{-1}(0) \right\} = 
         \int Dz
  \exp\left\{p_0\,G_0(\sqrt{\lambda_0}\, z)
      \right\}.
\end{equation}
In the limit $p_0 = n \to 0$ one recovers the usual expression \cite{Par80c}
\begin{equation}
\label{eq:sa-z000}
G_{-1}(0) =  \int Dz\, G_0(\sqrt{\lambda_0}\, z).
\end{equation}

Equation (\ref{eq:sa-gr}) has an interesting ``physical'' 
interpretation.
The quantity $G_R(b^R)$ is the free energy of a system of {\sl one} spin,
i.e. of size $1$, in the replica space
in presence of the frozen field $b^R$, that is with {\sl all} random 
(Gaussian) $z^r$ held fixed. 
To move one level up, $R\to R-1$, we have to unfreeze and 
integrate over $z^R$, while keeping all other fields $z^t$ with $t<R$ frozen.
The fields $z^t$ with $t<R$ give the effective action, 
under the form  of a (random) field, of the spins $s^b$ on the 
spin $s^a$ with $a\cap b = t < R$. Then integration over the field 
$z^R$ means that only the spins $s^a$ and $s^b$ such 
that $a\cap b = R$ are summed in the trace. 
All others are kept frozen.
Thus the quantity $G_{R-1}(b^{R-1})$ can be seen as the free energy (density) 
of a system in the replica space of {\sl size} $p_R$ in presence of
an external field $b^{R-1}$, which gives the 
interaction with the frozen spins, that is the 
{\sl frozen degrees of freedom}.
Extension to the successive $z^r$-integration is straightforward.
The quantity $G_{r-1}(b^{r-1})$ is obtained by integrating out in turn the  
random fields $z^t$ with $t\geq r$, while keeping all $z^t$ with $t<r$ frozen.
This means that the trace is restricted to spins $s^a$ and $s^b$ such that 
$a\cap b = t \geq r$. 
The contribution from the spins not included into the trace, and hence frozen, 
is taken into account by the frozen field $b^{r-1}$. 
The quantity $G_{r-1}(b^{r-1})$ is then the free energy (density) of a system 
of size $p_r$ 
in the replica space
in presence of the external field $b^{r-1}$, which accounts 
for the degrees of freedom still frozen at scale $r$.

The free energy $G_{-1}(0)$ is part of the total free energy density of the 
system, see eqs. (\ref{eq:e5}) and (\ref{eq:sa-gf}),
and thus it is itself an intensive quantity in the real space.
This implies that $p_r G_{r-1}$ are intensive quantities, and hence
as $n\to 0$ the $p_r$ become densities in the real space: 
$0 < p_r <1$.
The $p_r$ give a measure of the density of the frozen degrees of freedom 
at scale $r-1$ as measured from the overlap. Consider indeed the function
\begin{equation}
  x(q) = n + \sum_{r=0}^{R}(p_{r+1}-p_r)\,\theta(q - Q_r)
\end{equation}
which equals the number of pairs of replicas with 
overlap $Q^{ab}$ less or equal to $q$: $x(q) = p_{r+1}$ if $Q_r < q < Q_{r+1}$.
The function $x(q)$ is not decreasing with $q$, thus 
$p_r < p_{r'}$ if $r < r'$ as $n\to 0$.
Indeed in going from level $r$ to level $r-1$ the number of unfrozen degrees of
freedom, that is the number of spins in the replica space 
over which the trace is done, increases, and hence 
the number of {\sl frozen} degrees of freedom {\sl decreases},
as signaled by the decrease of the value of the overlap.
This picture is fully consistent with the dynamical formulation of 
CHS \cite{CriHorSom93,CriLeu07}
in terms of time-scales and density of frozen/unfrozen degrees of freedom.

We can now turn to the problem of calculating spin averages. 
This differs from that of ${\cal Z}(0)$ by the presence of terms
that depends on the fields $b^R_a$, cfr. 
eqs. (\ref{eq:sa-spaR}) and (\ref{eq:sa-z0}). 
The recursion relation (\ref{eq:sa-gr}) is the usual rule to compute
the free energy when some frozen degrees of freedom become unfrozen,
and hence must be summed up in the trace. In the specific case those frozen at
scale $r$ but unfrozen at scale $r-1$, represented by the fields $z^r$.
The presence of $p_r$ instead of $p_{r+1}$ 
in the integrand follows because at scale $r$ there are 
$p_r / p_{r+1}$ disjoint systems in the replica space, 
all with the same free energy, that merge at scale $r-1$.
This suggests the following recursion relation for the calculation of
spin averages.
Let $F_s(b^s_a)$ be a generic function of the field $b^s_a$ at scale $s$.
We then define the quantity $F_{s-1}(b^{s-1}_a)$ at scale $s-1$ 
as the average of $F_s(b^s_a)$ over the random field $z^s$ {\sl weighted} 
with the statistical weight of the state, that is,
\begin{eqnarray}
\label{eq:sa-Fs}
F_{s-1}(b^{s-1}_a) &=&
  Z_{s-1}(a)^{-1}
\nonumber\\
&\phantom{=}&
\times
         \int Dz^s_a\,
         Z_s(a)^{p_s/p_{s+1}}\, 
         F_s(\sqrt{\Delta\lambda_s}\, z^s_a + b^{s-1}_a).
\end{eqnarray}
%
%
where 
\begin{eqnarray}
      Z_r(\alpha) \equiv Z_r(b^r_\alpha)
      &=& \exp \bigl\{ p_{r+1}\,G_r(b^r_\alpha)\bigr\}
\nonumber\\
&=& 
  \exp\bigl\{p_{r+1}\,G_r(\sqrt{\Delta\lambda_r}\, z^r_\alpha
               + b^{r-1}_\alpha)\bigr\}
\end{eqnarray}
This recursion relation is supplemented by 
the boundary condition
\begin{equation}
F_s(b^s_a)\Bigr|_{s=r} = F_r(b^r_a)
\end{equation}
where $F_r(b_r)$ is a known expression.
Assume for example that
\begin{equation}
 F_r(b^r_a) = \frac{\partial}{\partial b^r_a}\,G_r(b^r_a)
\end{equation}
then a simple calculation shows that 
\begin{equation}
 F_{r-1}(b^{r-1}_a) = \frac{\partial}
                       {\partial b^{r-1}_a}\,G_{r-1}(b^{r-1}_a).
\end{equation}
This result is not unexpected since it just states that the magnetization 
$m_r$ at any scale $r$ is given by 
the derivative of the free energy of that scale with respect to the 
applied field at that scale: 
\begin{equation}
\label{eq:sa-mr}
  m_r(b^r_a) = \frac{\partial}{\partial b^r_a}\,G_r(b^r_a).
\end{equation} 
From this result it immediately follows that 
\begin{eqnarray}
\langle s^a\rangle &=& m_{-1}(h) 
                   = \frac{\partial}{\partial h} G_{-1}(h)
\nonumber\\
                   &=& \frac{\partial}{\partial h} 
                    \int Dz\, G_0(\sqrt{\lambda_0}\, z + h)
                   = \int Dz\, m_0(\sqrt{\lambda_0}\, z + h).     
\end{eqnarray}
where $h$ is an external field.%
\footnote{  
\label{note:bvsh}
  The field $h$ includes a factor $\beta$ from the
  statistical weight $\exp(-\beta H)$, see also 
  eq. (\protect\ref{eq:sa-gf}). Thus the correct expression 
  in terms of the real external field would have 
  $\beta h$. We prefer to leave the factor $\beta$ hidden 
  into the field to have no factors
  $\beta$ into
  the definition of the 
  spin averages via eq. (\protect\ref{eq:sa-spa}).
}
Clearly $\langle s^a\rangle = 0$ if $h=0$.

\begin{figure}
\includegraphics{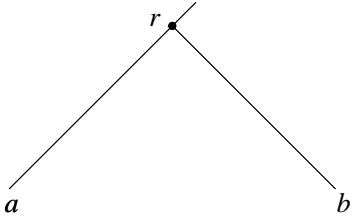}
\caption{$\langle s^a s^b\rangle_{a\cap b=r} = Q_r$ tree configuration.
        }
\label{fig:two-spin}
\end{figure}
To compute the two-spin correlation $\langle s^a s^b\rangle$ 
with $a\cap b = r$, i.e. the overlap $Q_r$, we have to evaluate the
integral
\begin{equation}
\langle s^as^b\rangle \rightarrow
           \int {\cal D}_R(\alpha) \prod_{0}^{R} 
                  Z_R(\alpha) \ 
                    m_R(b^R_a)\, m_R(b^R_b)
\end{equation}
where we used (\ref{eq:sa-mr}).
On a branch-tree diagram the two replicas 
$a$ and $b$ with $a\cap b = r$
are on different branches for scales $s > r$, see Fig. \ref{fig:two-spin},  
i.e., they belong to different systems.
The two local fields $b^s_a$ and $b^s_b$, and hence the
magnetizations, are independent from each 
other and the integration factorizes. Then
\begin{equation}
\label{eq:sa-m2i}
\langle s^a s^b\rangle \rightarrow
           \int {\cal D}_r(\alpha) \prod_{0}^{r} 
            Z_r(\alpha) \
                    m_r(b^r_a)^2
\end{equation}
since at scale $r$ the two replicas end up in the same system of size $p_r$
and the two fields $b^r_a$ and $b^r_b$ become equal.
This expression suggests introducing the
quantity 
\begin{equation}
\label{eq:sa-m2}
  F_s(b^s_a) = m_{r,s}^{(2)}\bigl(b^s_a)
\end{equation}
with the boundary condition
\begin{equation}
\label{eq:sa-m2bc}
 m_{r,r}^{(2)}(b^r_a) =  m_r(b^r_a)^2
\end{equation}
in terms of which we have
\begin{equation}
\label{eq:sa-2sp}
 \langle s^a s^b\rangle\Bigr|_{a\cap b = r} = m_{r,-1}^{(2)}(h)
    \stackrel{n\to 0}{=} 
       \int Dz\, m_{r,0}^{(2)}(\sqrt{\lambda_0}\, z + h)
\end{equation}
where $m_{r,0}^{(2)}$ is computed from the recursion relation (\ref{eq:sa-Fs})
with (\ref{eq:sa-m2}) and (\ref{eq:sa-m2bc}).

For higher order spin correlations we proceed in a similar way.
Consider for example the four-spin correlation 
\begin{figure}
\includegraphics{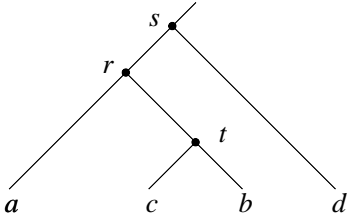}
\caption{Longitudinal-Anomalous Sector tree configuration.}
\label{fig:four-spinLA}
\end{figure}
%
\begin{equation}
\fl
\langle s^a s^b s^c s^d\rangle^{r,s}_{t} \rightarrow
           \int {\cal D}_R(\alpha) \prod_{0}^{R} 
           Z_R(\alpha) \
                    m_R(b^R_a)\, m_R(b^R_b)\,
                    m_R(b^R_c)\, m_R(b^R_d)
\end{equation}
where the replicas $a,b,c,d$ have the LA Sector configuration shown in 
Fig. \ref{fig:four-spinLA}.
The four replicas are independent from scale $R$ to $t$, 
where the replicas $b$ and $c$ end up in
the same system in the replica space 
and the fields $b^t_b$ and $b^t_c$ become equal.
Thus
\begin{equation}
\langle s^a s^b s^c s^d\rangle^{r,s}_{t} \rightarrow
           \int {\cal D}_t(\alpha) \prod_{0}^{t} 
             Z_t(\alpha) \
                    m_t(b^t_a)\, m_t(b^t_b)^2\,
                    m_t(b^t_d).
\end{equation}
Moving up along the tree
the surviving three replicas remain in different systems up to scale $r$,
where the replicas $a$ and $b$ eventually find themselves into the same 
system. Then, by using the quantity $m_{r,s}^{(2)}$ introduced for the 
two-spin correlation, we can write
\begin{equation}
\langle s^a s^b s^c s^d\rangle^{r,s}_{t} \rightarrow
           \int {\cal D}_r(\alpha) \prod_{0}^{r} 
            Z_r(\alpha) \
                    m_r(b^r_a)\, m_{t,r}^{(2)}(b^r_a)\,
                    m_r(b^r_d)
\end{equation}
where $m_{t,r}^{(2)}(b^r)$ is obtained from the recursion relation
(\ref{eq:sa-Fs}) with the initial condition 
$m_{t,t}^{(2)}(b^t) = m_t(b^t)^2$. For the next step we observe that
replicas $a$ and $d$ remain into different subspaces up to scale $s$. 
Thus by introducing the quantity 
\begin{equation}
F_k(b^k_a) = m_{t,r,k}^{(2,1)}(b^k_a)
\end{equation}
with the boundary condition,
\begin{equation}
m_{t,r,r}^{(2,1)}(b^r_a) =  m_r(b^r_a)\, m_{t,r}^{(2)}(b^r_a)
\end{equation}
we can move up along the tree up to scale $s$, and
\begin{equation}
\langle s^a s^b s^c s^d\rangle^{r,s}_{t} \rightarrow
           \int {\cal D}_s(\alpha) \prod_{0}^{s} 
           Z_s(\alpha) \
                    m_{t,r,s}^{(2,1)}(b^s_a)\,
                    m_s(b^s_a).
\end{equation}
The last step from scale $s$ to scale $0$ is now straightforward. 
We introduce the quantity
\begin{equation}
F_k(b^k_a) = m_{t,r,s,k}^{(2,1,1)}(b^k_a)
\end{equation}
with the boundary condition,
\begin{equation}
m_{t,r,s,s}^{(2,1,1)}l(b^s_a) = m_{t,r,s}^{(2,1)}(b^s_a)\, m_s(b^s_a)
\end{equation}
so that the final result reads
\begin{equation}
\langle s^a s^b s^c s^d\rangle^{r,s}_{t} =
           m_{t,r,s,-1}^{(2,1,1)}(h)  
    \stackrel{n\to 0}{=} 
   \int Dz\, m_{t,r,s,0}^{(2,1,1)}(\sqrt{\lambda_0}\, z + h).
\end{equation}

In a similar way, once the replica geometry is specified, one can compute 
spin averages involving any number of spins.

The above results, valid for any finite $R$, are easily extended to 
the continuous case
$R\to\infty$. In this case, since the values of $Q^{ab}$, and hence those of
$\lambda_r$, are bounded in a finite interval, the differences 
$\Delta\lambda_r\to 0$ as $R\to\infty$ to account for an infinite number of values 
in a finite interval.\footnote{One can allow for a finite number of 
``jumps'', that is points where $\Delta_r$ does not vanishes as $R\to\infty$.
One then gets mixed-type solutions as those found, e.g. in spherical 
$p$-spin models \cite{CriLeu06,CriLeu07b}.} 
As a consequence the recursion relations
are replaced by differential equations. In particular the recursion 
relation (\ref{eq:sa-gr}) becomes the Parisi equation \cite{Par80c}, while
eq. (\ref{eq:sa-Fs}) is replaced by the partial differential equation
\begin{equation}
\label{eq:sa-eqF}
 \dot{F}(x,b) = -\frac{\dot{\lambda}(x)}{2}
                       \Bigl[
                F''(x,b) + 2\,x\, m(x,b)\,F'(x,b)
                                        \Bigr], \qquad x < t
\end{equation}
where $m(x,b) = G'(x,b)$, 
with the initial condition 
\begin{equation}
\label{eq:sb-bcF}
  F(x,b)\Bigl|_{x=t} = F(t,b)
\end{equation}
where $F(t,b)$ is some known expression at scale $t$.
As usual the ``dot'' and the ``prime'' denote partial derivative with respect to
$x$ and $b$, respectively. Details are in \ref{app:SA-Cont}.

We conclude this Section by noticing that this formalisms can be easily 
extended to calculate the (static) response of the system to
external perturbations that act at given scales. Let us denote by 
$\epsilon^r_a$ a small external perturbation acting {\sl only} on scale $r$.
We then define the static response as
\begin{equation}
\label{eq:sa-chi}
 \chi_r = \frac{\partial}{\partial\epsilon^r_a}\langle 
                 s^a\rangle^{\epsilon^r}\Bigr|_{\epsilon^r=0}
\end{equation} 
where $\langle s^a\rangle^{\epsilon^r}$ is the spin average in presence 
of $\epsilon^r$. 
The perturbation $\epsilon^r_a$ can be seen as an extra contribution to
the frozen field $b^r_a$. Then, since $\epsilon^r_a$ acts {\sl only} 
on scale $r$, in the recursion relation we end up with
$G_{r+1}(\sqrt{\Delta\lambda_{r+1}}z^{r+1}_a + b^r_a + \epsilon^r_a)$.
By expanding to the first order in $\epsilon^r_a$, we finally have
\begin{eqnarray}
\fl
\label{eq:sa-seps}
\langle s^a \rangle^{\epsilon^r} &\rightarrow&
           \int {\cal D}_{r}(\alpha) 
    \prod_{(\alpha_0,\ldots,\alpha_r)\not=(a_0,\ldots,a_r)}
    \int\,Dz^{r+1}_\alpha\,
                  \exp \left\{ p_{r+1}G_{r+1}(b^{r+1}_\alpha)\right\} 
\nonumber\\
\fl
&\phantom{.}& 
  \times
    \int\,Dz^{r+1}_a
                    \exp \left\{ p_{r+1}G_{r+1}(b^{r+1}_a) + 
                                 p_{r+1}\epsilon^r_a m_{r+1}(b^{r+1}_a)
                         \right\} \,
                    m_{r+1}(b^{r+1}_a)
\end{eqnarray}
and a similar expression for the normalization ${\cal Z}_{\epsilon^r}(0)$.
By neglecting all unnecessary indexes,
the recursion relation (\ref{eq:sa-Fs}) is then replaced by 
\begin{eqnarray}
\fl
 m_r^{\epsilon^r}(b) &=& 
          \exp \left\{-p_{r+1}\, G_r^{\,\epsilon^r}(b)\right\}
\nonumber\\
\fl
 &\phantom{=}& 
 \times
               \int Dz 
               \exp\left\{ p_{r+1}G_{r+1}(\sqrt{\Delta\lambda_{r+1}}\,z+b)
              + p_{r+1}\,\epsilon^r\,m_{r+1}(\sqrt{\Delta\lambda_{r+1}}\,z+b)
                   \right\}\
\nonumber\\
\fl
 &\phantom{=}& 
    \phantom{xxxxxxxxxxxxxxxxxxxxxxxx} \times
                 m_{r+1}(\sqrt{\Delta\lambda_{r+1}}\,z+b)
\end{eqnarray}
with
\begin{eqnarray}
\fl
          \exp \left\{p_{r+1}\, G_r^{\,\epsilon^r}(b)\right\} &=&
               \int Dz 
               \exp\Bigl\{ p_{r+1}G_{r+1}(\sqrt{\Delta\lambda_{r+1}}\,z+b)
\nonumber\\
&\phantom{=}&
\phantom{=============}
              + p_{r+1}\,\epsilon^r\,m_{r+1}(\sqrt{\Delta\lambda_{r+1}}\,z+b)
                   \Bigr\}.
\end{eqnarray}
Taking the derivative with respect to $\epsilon^r$, and setting 
$\epsilon^r\to 0$, leads to
\begin{equation}
\frac{\partial}{\partial\epsilon^r_a}\, m_r^{\epsilon^r}(b)\Bigr|_{\epsilon^r=0}
     = p_{r+1} \left(  m_{r+1,r}^{(2)}(b) - m_r(b)^2 \right).
\end{equation}
Then from eqs. (\ref{eq:sa-chi}) and (\ref{eq:sa-seps}) we have%
\footnote{  
  The perturbation $\epsilon^r_a$ includes a factor $\beta$. This removes the
  factor $\beta$ on the r.h.s.
}
\begin{equation}
 \label{eq:sa-chir}
  \chi_{r} = p_{r+1}\,\left(Q_{r+1} - Q_r\right).
\end{equation}
The derivative of the normalization factor ${\cal Z}_{\epsilon^r}(0)$ gives
a term proportional to $\langle s^a\rangle^2$, which vanishes in absence
of external field. The expression (\ref{eq:sa-chir}) is the static limit
of the modified Fluctuation Dissipation Theorem introduced by 
CHS \cite{CriHorSom93,CriLeu07}
in dynamics.


\section{Replicon Sector}
\label{sec:Repl}
\begin{figure}
\includegraphics{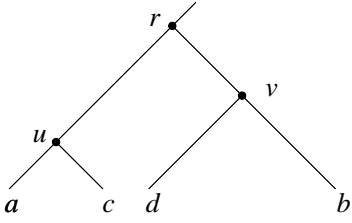}
\caption{Replicon Sector tree configuration.}
\label{fig:four-spinR}
\end{figure}

The Hessian is a $\frac{n(n-1)}{2}\times\frac{n(n-1)}{2}$ symmetric matrix 
that after block-diagonalization \cite{DeDomCarTem97,TemDeDomKon94} 
becomes a string of
$(R+1)\times(R+1)$ blocks along the diagonal for the LA Sector, followed by 
$1\times 1$ fully diagonalized blocks, for the Replicon Sector.
The diagonal elements in the Replicon Sector are given by:
\begin{equation}
\label{eq:e25}
\lambda_{\rm R}(r;\hat{k}\hat{l}) = M_{\hat{k};\hat{l}}^{r;r}.
\end{equation}
To evaluate $M_{\hat{k};\hat{l}}^{r;r}$ we need the matrix elements
$M_{u;v}^{r;r}$, eq. (\ref{eq:e9}), that is the four-spin average 
$\langle s^a s^b s^c s^d\rangle$ for the Replicon Geometry shown in 
Figure \ref{fig:four-spinR}.
From the results of Section \ref{sec:SA} and \ref{app:SA-Cont} 
this is given,
in absence of an applied external field and for $R\to\infty$,
by
\begin{equation}
\langle s^a s^b s^c s^d\rangle_{u;v}^{r;r} = m_{u,v}^{(2,2)}(0,0).
\end{equation}
The function $m_{u,v}^{(2,2)}(x,b)$ is solution of a chain of
partial differential equation of the form
\begin{equation}
\label{eq:eqF}
 \dot{F}(x,b) = -\frac{\beta^2\dot{Q}(x)}{2}
                       \Bigl[
                F''(x,b) + 2\,x\, m(x,b)\,F'(x,b)
                                        \Bigr],
\end{equation}
where $m(x,b)$ is the local magnetization at scale $x$ in 
presence of the field $b$.

In our case, starting from the bottom of the tree in 
Fig. \ref{fig:four-spinR}, we first have to solve eq. (\ref{eq:eqF}) for
$F(x,b) = m(x,b)$ and boundary condition
\begin{equation}
 F(x,b)\Bigl|_{x=1^-} = m(1,b) = \tanh b.
\end{equation}
The range of $x$ is $u\leq x\leq 1$ for the left branch of the tree, and
$v\leq x\leq 1$ for the right branch.
To proceed towards scale $0$ we have to solve next eq. (\ref{eq:eqF}) for
$F(x,b) = m_{t}^{(2)}(x,b)$ and initial condition
\begin{equation}
  F(x,b)\Bigl|_{x=t} = m_{t}^{(2)}(t,b) = m(t,b)^2
\end{equation}
where $t=u,v$ depending upon we are on the left or on the right branch of the tree.
The range of $x$ is either $r\leq x\leq u$, left branch, or
$r\leq x\leq v$, right branch, see Fig. \ref{fig:four-spinR}.

To accomplish the last step, $0\leq x\leq r$, we finally solve equation
(\ref{eq:eqF}) for $F(x,b) = m_{u,v}^{(2,2)}(x,b)$ and 
initial condition
\begin{equation}
\label{eq:bcF}
  F(x,b)\Bigl|_{x=r} = m_{u,v}^{(2,2)}(r,b)
                     = m_{u}^{(2)}(r,b)\, m_{v}^{(2)}(r,b).
\end{equation}

While these equations are valid for any temperature $T$, we are interested into 
their solution in the limit $T\to 0$. Following Pankov \cite{Pan06}
one can show, see \ref{app:Pankov}, that
in the {\sl outer region} $x \gg T$ and $T\to 0$ the solution of the partial 
differential equation (\ref{eq:eqF}) looses its explicit dependence on the
scale variable $x$. As a consequence for $R\to\infty$ 
the matrix element $M_{u;v}^{r,r}$ becomes independent of $u$ and $v$ for 
all $r$ such that $p_r = x(Q_r) \gg T$ as $T\to 0$, and this in 
turns implies that $M_{\hat{k};\hat{l}}^{r;r}$ is independent of $k$ and $l$.
Thus by exploiting this {\sl insensitivity} we conclude that
for all $r$ in the {\sl outer region}
\begin{equation}
M_{\hat{k};\hat{l}}^{r;r} =  M_{\widehat{r+1};\widehat{r+1}}^{r;r}
                          = O\left(\frac{1}{R^2}\right) 
                  \stackrel{R\to\infty}{=} 0.
\end{equation}
The second equality follows from a Ward-Takahashi identity 
\cite{DeDomKonTem98}.
In the inner region the Replicon spectrum maintains its complexity. However
its relevance becomes less and less important as $T$ approaches zero,
and vanishes in the limit $T\to 0$ when the thickness of the boundary 
shrinks to zero. 
The Replicon spectrum, similarly to the order parameter function $Q(x)$, 
becomes then discontinuous at $x=0$.

\section{Longitudinal-Anomalous Sector}
\label{sec:LA}
The LA Sector corresponds to the $(R+1)\times(R+1)$ 
diagonal blocks along the diagonal labeled by the index 
$k=0,\ldots, R+1$. The matrix element in each block turns out to be
\begin{equation}
\label{eq:LAs}
 \leftidx{_{\rm LA}}M_{\hat{k}}^{r;s} = 
     \Lambda_{\hat{k}}(r)\,\delta_{r,s}^{\rm Kr}
     + \frac{1}{4} M_{\hat{k}}^{r;s}\,\delta_s^{(k)}, 
\qquad r,s = 0, \ldots, R
\end{equation}
where $\Lambda_{\hat{k}}(r)$ is a shorthand for
\begin{equation}
  \Lambda_{\hat{k}}(r) = \left\{\begin{array}{ll}
     M_{\hat{k};\widehat{r+1}}^{r;r} & k > r+1, \\
     \phantom{x} & \phantom{x} \\
     M_{\widehat{r+1};\widehat{r+1}}^{r;r} & k \leq r+1, \\
\end{array}\right.
\end{equation}
and $\delta_{s}^{(k)} = p_s^{(k)} - p_{s+1}^{(k)}$, $k=0,1,\ldots,R+1$, 
with
\begin{equation}
\label{eq:e35}
p_s^{(k)} = \left\{\begin{array}{ll}
  p_s & s\leq k \\
2 p_s & s > k.
\end{array}\right.
\end{equation}
$M_{\hat{k}}^{r;s}$ is the RFT of the matrix element 
$M_{t}^{r;s}$ with respect the cross-overlap $t$, that is,
\begin{equation}
\label{eq:e29}
 M_{\hat{k}}^{r;s} = \sum_{t=k}^{R+1} p_t^{(r,s)}
               \left(M_t^{r;s} - M_{t-1}^{r;s}\right)
\end{equation}
with , if $r<s$,  
\begin{equation}
p_t^{(r,s)} = \left\{\begin{array}{ll}
  p_t & t\leq r \\
2 p_t & r < t \leq s \\
4 p_t & r < s < t
\end{array}\right.
\end{equation}
For finite temperature the matrix elements are different in each block, 
however for scales $k$ in the outer region
the RFT $M_{\hat{k}}^{r;s}$ and $\Lambda_{\hat{k}}(r)$ become
insensitive to the value of $k$ as $T\to 0$. All correspondent 
blocks are then diagonalized 
through the single eigenvalue equation:\footnote{
  The boundary term $t=0$ in the RFT is proportional to $p_0 =n$ 
  and vanishes for $n\to 0$. The next
  term is proportional to $p_1\,M_{1}^{r;s}$, since 
  $M_{0}^{r;s} = 0$, and  vanishes as $R\to\infty$,
  so the only term which survives is $k=R+1$.
}
\begin{equation}
\label{eq:e28}
\lambda_{\rm LA} f^r = M_{\widehat{R+1};\widehat{r+1}}^{r;r} f^r 
             +\frac{1}{4}\sum_{s=0}^{R} M_{\widehat{R+1}}^{r;s} \delta_s f^s
\end{equation}
where $\delta_s = p_s - p_{s+1}$.
In the outer region the eigenvectors $f^r$ satisfy 
$f^r\not=0$ if $T\ll x(Q_r) \leq x_c$ as $T\to 0$, and zero otherwise.
The eigenvalue equation then reduces becomes
\begin{equation}
\label{eq:e30}
 \lambda_{\rm LA} f^r = \frac{1}{4} M_{\widehat{R+1}}^{R;R} 
 \sum_{s=\overline{r}}^{R} \delta_s f^s,
\qquad r = \overline{r},\ldots, R.
\end{equation}
where $\overline{r}$ is the lower bound of the outer region: 
$x(Q_{\overline{r}}) = \overline{x} \sim \delta$
as $T\to 0$.
The diagonal Replicon contribution vanishes for
$R\to\infty$, as ensured by the Ward-Takahashi 
identity, and does not contribute. This equation has two distinct solutions. 
The first
\begin{equation}
\label{eq:la0}
\lambda_{\rm LA} = 0
\end{equation} 
for $\sum_{s=\overline{r}}^{R}\delta_s f^s=0$,
and
\begin{eqnarray}
\label{eq:la1}
\lambda_{\rm L} &=& \frac{1}{4}\left(\sum_{s=\overline{r}}^{R}\delta_s\right)\, 
            M_{\widehat{R+1}}^{R;R} 
\nonumber\\
            &=&
            (\overline{x} - 1) 
            \bigl(1 - \beta^2\,(1-q_c(T)\bigr)
\nonumber\\
            &=&
            (\alpha - 1) + O(T), 
\quad T\to 0
\end{eqnarray}
for $\sum_{s=\overline{r}}^{R}\delta_s f^s\not=0$.
The last equality follows from eq. (\ref{eq:qc}), and
$\overline{x} \sim \delta \sim T$
as $T\to 0$.
In the inner region, where the LA spectrum maintains the RSB structure,
the solutions are smooth functions of the inner variable even for $T\to 0$.
For $T\to 0$ the thickness of the boundary layer shrinks to zero, and
the eigenvalues (\ref{eq:la0}) and (\ref{eq:la1}) cover the whole
LA spectrum, with a discontinuity at $x=0$.

\begin{figure}
\includegraphics{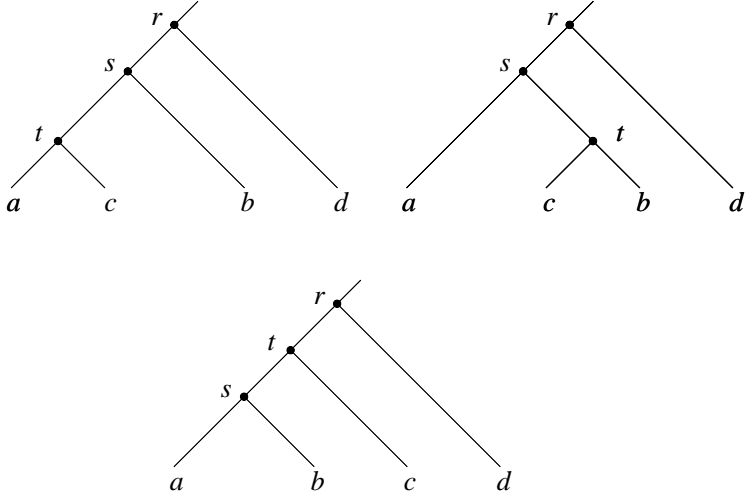}
\caption{Possible tree configurations corresponding to the 
Longitudinal-Anomalous Sector with $r<s$ and $t\geq r$.
}
\label{fig:four-spinLAf}
\end{figure}

\section{Summary and Conclusions}
\label{sec:disc}

In this work we have presented the analysis of the very low temperature limit
$T\ll 1$ of the spectrum of the Hessian 
for the Parisi solution of the SK model. 
It has been long known that in this regime two distinct 
regions of the interval $x\in[0,1]$ can be identified according to the 
variation of the order parameter function $Q(x)$ with $x$. 
We have shown that this has strong consequences on the 
the structure of the Hessian spectrum. 
In the first region $x \leq \delta \sim T$, where $Q(x)$ varies rapidly 
from $Q(0)=0$ up to $Q(\delta) \sim q_c(T)\sim 1$, the spectrum maintains the 
complex structure observed close to the critical temperature $T_c$ 
for the full RSB state. We can call this region the {\sl RSB-like} regime.
In the second region, $T \ll x \leq x_c$ with $x_c\sim 0.575\ldots$,
where $Q(x)$ is slowly varying, the Hessian spectrum has a completely 
different structure. Here the components of the
Hessian matrix become insensitive to changes of the overlaps and the
bands observed in the replica symmetry breaking regime
collapse. In this region only two distinct eigenvalues survive: 
a null one and the positive one. This ensures that 
the Parisi solution of the SK model then remains stable as the temperature 
goes to zero. Remarkable is the occurrence of zero modes in 
both the Replicon and the LA Sectors. 
Null eigenvalues arise from Replicon geometry, with 
Ward-Takahashi identities protecting them. Note, however, that 
the zero modes arise also from LA geometry, that is without protection of 
the Ward-Takahashi identities.

We observe that for $T\ll 1$ the order parameter function is almost 
constant for $T\ll x\leq x_c$, the variation being indeed of order 
$[Q(x_c)-Q(x)]/Q(x) = {\cal O}\left((T/x)^2\right)$. Thus in this region
we have a marginally stable (almost) replica symmetric solution, that becomes 
a genuine replica symmetric solution in the limit $T\to 0$, with self-averaging 
trivially restored.
It is worth to remind 
that the stability analysis of the replica symmetric solution
also leads to two eigenvalues, one of which is zero to the lowest order
in $T_c-T$ (and negative to higher order), and the other positive.

In the limit $T\to 0$ the region where the RSB structure of the solution 
is found shrinks to zero, and only the RS part survives. 
This feature, in a sense, brings about some perfume of conciliation between 
aspects of Parisi mean-field approach and of the droplet approach
\cite{AspMooYou03,CrideDom10}.
We stress, however, that in order to identify a genuine droplet
behavior, corrections to the mean-field have to be studied in more details.

Concerning the multiplicity of the eigenvalues we observe that 
in each Sector, Replicon and LA, one has to separate the contribution from 
the RSB-like and the droplet-like regions. The former is proportional to
the width $\delta$ of the region. Therefore in the limit $T\to 0$ the
contribution from the RSB-like region vanishes, and one has the usual
Replicon and LA multiplicities for the droplet-like region.

Finally, we have developed a method to compute spin averages 
in replica space involving any number of replicas, both for a finite
number $R$ of replica symmetry breaking steps and for the continuous
limit $R\to\infty$. This generalizes some special cases known for 
the continuous limit $R\to\infty$, and to our knowledge it is new.
Moreover it sheds light on the interpretation of the replica symmetry breaking 
method and its relation with the dynamical approach. For example we were
able to compute the static susceptibility and show that it equals 
the static limit of the dynamic susceptibility computed via the 
modified Fluctuation Dissipation Theorem.

\ack
The authors acknowledge useful discussions with 
M. A. Moore, 
R. Oppermann, T. Sarlat and A. P. Young.
A.C. acknowledges hospitality and support from 
IPhT of CEA, where part of this work was done.

\appendix

\section{Spin Averages: continuous case}
\label{app:SA-Cont}
The expressions derived in Section \ref{sec:SA} are valid for any finite
number $R$ of replica symmetry breaking steps.
Here we shall address the limit $R\to\infty$, where the replica symmetry
breaking become continuous.
If the values of $\lambda_r$ are bounded in a finite interval,
as is the case of the SK model, in the limit $R\to\infty$
they must be dense and 
\begin{equation}
 \Delta\lambda_r \to 0 \
\mbox{\rm as}\ R\to\infty, 
 \qquad r = 0, 1,\ldots, R.
\end{equation}
so that the integral in the recursion relations can then be evaluated  
by expanding in powers of $\Delta\lambda_r$.
We observe that a finite number of ``jumps'', that is 
values of $r$ where $\Delta\lambda_r$ remains finite as $R\to\infty$, 
are possible. One then gets mixed continuous-discrete phase
described by a piecewise order parameter function,   
as those observed, e.g., in spherical $p$ spin-glass models 
\cite{CriLeu06,CriLeu07b}.
We shall not discuss this case here.

Let us first consider the recursion relation for the free energy $G_r$. 
In the following we drop all unnecessary indexes. 
By expanding $G_r(\sqrt{\Delta\lambda_r}\, z + b)$ in eq. (\ref{eq:sa-gr})
in powers of $\Delta\lambda_r$
and integrating over the Gaussian variable $z$, a straightforward algebra
leads to
\begin{equation}
\label{eq:b-recG}
 G_{r-1}(b) = G_r(b) 
            + \frac{\Delta\lambda_r}{2} \left[
                G_r''(b) + p_r \bigl(G_r'(b)\bigr)^2
                                        \right]  + O(\Delta\lambda_r^2)
\end{equation}
where the ``prime'' denote differentiation with respect the argument, i.e. the 
field $b$:
\begin{equation}
  G_r'(b) = \frac{\partial}{\partial b} G_r(b).
\end{equation}
%
Next we observe that $G_r(b)$ is a function of $p_r$, thus to extract the 
non-trivial part of eq. (\ref{eq:b-recG}) as $\Delta\lambda_r\to 0$ we have
to specify what happens to $p_r - p_{r-1}$ as $R\to\infty$. 

Suppose $p_r - p_{r-1}$ does not vanish in the limit $R\to\infty$. 
In this case eq. (\ref{eq:b-recG}) implies that
\begin{equation}
\label{eq:b-plat}
 G_{r-1}(b) = G_r(b) \qquad \mbox{as}\ \Delta\lambda_r\to 0 \
                            \mbox{but}\ p_r-p_{r-1} \not= 0
\end{equation}
that is, $G_r(b)$ does not depend on the scale. This is what happens, 
for example, in the SK model for $x_c < r < 1$, the so called ``plateau''.
If $p_r - p_{r-1} \to 0$ as $\Delta\lambda_r\to 0$
then, by defining 
\begin{equation}
  p_r \equiv x \qquad \Rightarrow\qquad  p_r - p_{r-1} = \Delta x
\end{equation}
and changing the notation to $G_r(b) = G(x,b)$, equation 
(\ref{eq:b-recG}) leads for $\Delta\lambda_r\to 0$ to the Parisi equation
\cite{Par80c,Dup81}
\begin{equation}
\label{eq:b-parisi}
 \dot{G}(x,b) = -\frac{\dot{\lambda}(x)}{2}
                       \left[
                G''(x,b) + x \bigl(G'(x,b)\bigr)^2
                                        \right]
\end{equation}
where the ``dot'' denotes the (left) derivative with respect to 
the scale $x$, e.g.,
\begin{equation}
  \dot{G}(x,b) = \frac{\partial}{\partial x} G(x,b) 
               \equiv \lim_{\Delta x\to 0} \frac{G(x,b) - G(x-\Delta x,b)}
                                                {\Delta x}.
\end{equation}
We note that with this definition the validity of the
partial differential equation (\ref{eq:b-parisi}) can be extended to include
the case (\ref{eq:b-plat}) since the derivative is always well 
defined.
As a consequence the initial condition simply reads, see eq. (\ref{eq:sa-GR}),
\begin{equation}
 G(x=1^{-}, b) = \ln \bigl(2\,\cosh b\bigr)
\end{equation}
while (\ref{eq:sa-z000}) becomes
\begin{equation}
 G_{-1}(b) = \int Dz\, G(0,\sqrt{\lambda(0)}\, z + b).
\end{equation}
For the recursion relation (\ref{eq:sa-Fs}) we follows a similar procedure,
that is we expand the r.h.s. 
of (\ref{eq:sa-Fs}) in powers of $\Delta\lambda_r$ and integrate over the 
Gaussian variable $z$. This leads to
\begin{equation}
F_{r-1}(b) = F_r(b) + \frac{\Delta\lambda_r}{2} \Bigl[
              F_{r}''(b) + 2 p_r G_r'(b) F_r'(b)
              \Bigr] + O(\Delta\lambda_r^2).
\end{equation}
As before if $p_r-p_{r-1}$ does not vanish as $\Delta\lambda_r\to 0$, 
the recursion relation reduces to
\begin{equation}
  F_{r-1} (b) = F_r(b).
\end{equation}
If, however, $p_r - p_{r-1} = \Delta x \to 0$ as $\Delta\lambda_r\to 0$
we end up with the partial differential equation
\begin{equation}
\label{eq:b-eqF}
 \dot{F}(x,b) = -\frac{\dot{\lambda}(x)}{2}
                       \Bigl[
                F''(x,b) + 2\,x\, m(x,b)\,F'(x,b)
                                        \Bigr], \qquad x < t
\end{equation}
where $m(x,b) = G'(x,b)$ is the magnetization at scale $x$ in presence of
a field $b$, 
with the initial condition
\begin{equation}
\label{eq:b-bcF}
  F(x,b)\Bigl|_{x=t} = F(t,b)
\end{equation}
where $F(t,b)$ is a known expression at scale $t$.

As an example consider the two-spin correlation 
$\langle s^a s^b\rangle$ with $a\cap b = r$. From eq. (\ref{eq:sa-2sp})
it readily follows that 
\begin{equation}
\label{eq:b-2sp}
 \langle s^a s^b\rangle\Bigr|_{a\cap b = r} = 
   \int Dz\, m_{r}^{(2)}(0,\sqrt{\lambda(0)}\, z + h)
\end{equation}
where $m_r^{(2)}(x,b)$ is solution of the the partial differential 
equation
\begin{equation}
\label{eq:b-eqm2}
 \dot{m}_r^{(2)}(x,b) = -\frac{\dot{\lambda}(x)}{2}
                       \Bigl[
                {m_r^{(2)}}''(x,b) + 
                2\,x\, m(x,b)\, {m_r^{(2)}}'(x,b)
                                        \Bigr]
\end{equation}
for $0 < x < r$ and 
initial condition
\begin{equation}
 m_r^{(2)}(x=r,b) = m(r,b)^2.
\end{equation}
To our knowledge this equation was first derived by Goltsev \cite{Gol84}.

For the four-spin correlation $\langle s^a s^b s^c s^d\rangle^{r,s}_t$ of 
Fig. \ref{fig:four-spinLA} we have a similar expression:
\begin{equation}
\langle s^a s^b s^c s^d\rangle^{r,s}_{t} =
   \int Dz\, m_{t,r,s}^{(2,1,1)}(0,\sqrt{\lambda(0)}\, z + h)
\end{equation}
where $m_{t,r,s}^{(2,1,1)}(x,b)$ is solution of the partial differential 
equation
\begin{equation}
\fl
\label{eq:b-eqm211}
 \dot{m}_{t,r,s}^{(2,1,1)}(x,b) = -\frac{\dot{\lambda}(x)}{2}
                       \Bigl[
                {m_{t,r,s}^{(2,1,1)}}''(x,b) + 
                2\,x\, m(x,b)\, {m_{t,r,s}^{(2,1,1)}}'(x,b)
                                        \Bigr]
\end{equation}
for $0 < x < s$ and
initial condition
\begin{equation}
 m_{t,r,s}^{(2,1,1)}(x=s,b) = m_{t,r}^{(2,1)}(s,b)\, m(s,b).
\end{equation}
The function 
$m_{t,r}^{(2,1)}(x,b)$ is itself solution of the partial differential 
equation
\begin{equation}
\label{eq:b-eqm21}
 \dot{m}_{t,r}^{(2,1)}(x,b) = -\frac{\dot{\lambda}(x)}{2}
                       \Bigl[
                {m_{t,r}^{(2,1)}}''(x,b) + 
                2\,x\, m(x,b)\, {m_{t,r}^{(2,1)}}'(x,b)
                                        \Bigr]
\end{equation}
for $s < x < r$
and initial condition
\begin{equation}
 m_{t,r}^{(2,1)}(x=r,b) = m_{t}^{(2)}(r,b)\, m(r,b).
\end{equation}
Finally $m_t^{(2)}(x,b)$ is solution of equation (\ref{eq:b-eqm2}) for 
$r < x < t$ and initial condition $m_t^{(2)}(x=t,b) = m(t,b)^2$.

We conclude this Section by noticing that if we take $F(x,b) = m(x,b)$, then
the differential equation (\ref{eq:b-eqF}) becomes the known differential 
equation\cite{SomDup84},
\begin{equation}
\label{eq:b-eqm}
 \dot{m}(x,b) = -\frac{\dot{\lambda}(x)}{2}
                       \Bigl[
                m''(x,b) + 2\,x\, m(x,b)\,m'(x,b)
                                        \Bigr]
\end{equation}
with the initial condition
\begin{equation}
 m(1^{-},b) = \tanh b
\end{equation}
derived by Sommers and Dupont for the local magnetization.
Numerical solution of these equations can be found by using, e.g., the
method described in Ref. \cite{CriRiz02}.

\section{The Pankov scaling regime}
\label{app:Pankov}
To discuss the Pankov regime $T\to 0$ and $ T \ll x\ll 1$
we first perform the change of variable
$b = \beta y$ into the partial differential equation 
(\ref{eq:b-eqF}) to make explicit the temperature dependence. 
In the new variable the equation reads
\begin{equation}
\label{eq:c-eqF}
 \dot{F}(x,y) = -\frac{\dot{\lambda}(x)}{2\beta^2}
                       \Bigl[
                F''(x,y) + 2 \beta x\, m(x,y)\,F'(x,y)
                                        \Bigr]
\end{equation}
where the ``prime'' now denotes differentiation with respect to $y$, and
the local magnetization is
$m(x,y) = \beta^{-1}\,G'(x,y)$, see footnote page \pageref{note:bvsh}.
Following Pankov \cite{Pan06} we assume that 
the dependence on the local fields $y$ is via the combination 
$z=\beta x y$, that is,
\begin{equation}
\label{eq:c-pan}
  F(x,y) = \widetilde{F}(x,z), \qquad z = \beta x y
\end{equation}
and similarly for $m(x,y)$.
The differential equation (\ref{eq:c-eqF}) then becomes
\begin{eqnarray}
\label{eq:c-eqFz}
  x\,\dot{\widetilde{F}}(x,z) &=& -\frac{x^3\,\dot{\lambda}(x)}{2}
             \Bigl[ \widetilde{F}''(x,z) + 
                   2 \widetilde{m}(x,z)\, \widetilde{F}'(x,z)
             \Bigr] 
\nonumber\\
&\phantom{=}&
\phantom{===}
           -z \widetilde{F}'(x,z).
\end{eqnarray}
Pankov has shown that in the outer region the ``tilded'' functions
do not depend explicitly on the scale variable $x$: 
$\widetilde{F}(x,z) = \widetilde{F}(z)$. 
All dependence on scale, field and temperature enters via 
the combination $\beta x y$. Pankov called this the {\sl scaling regime}.

From eq. (\ref{eq:c-eqFz}) it is clear that the Pankov scaling regime is
only possible iff
\begin{equation}
\label{eq:c-psr}
 \frac{x^3\,\dot{\lambda}(x)}{2} = c = \mbox{\rm constant}
\quad \Rightarrow\quad
 \lambda(x) = \mbox{\sl const.} - \frac{c}{x^2}
\end{equation}
in which case the partial differential equation (\ref{eq:c-eqFz}) reduces
to the ordinary second order differential equation
\begin{equation}
\label{eq:c-eqFp}
\widetilde{F}''(z) = -\left[\frac{z}{c} + 2\,\widetilde{m}(z)\right]\,
                          \widetilde{F}'(z).
\end{equation}
In the SK model $\lambda(x) = \beta^2\,Q(x)$, where $Q(x)$ is the 
order parameter function, then from
(\ref{eq:c-psr}) it follows that in the  outer region $Q(x)$ has the form
\begin{equation}
 Q(x) = \mbox{\sl const.} - \frac{c}{(\beta x)^2}.
\end{equation}
Equation (\ref{eq:q-pank}) now follows by imposing 
$Q(x_c) = q_c(T) = 1 - \alpha T^2 + {\cal O}(T^3)$. 

\section{Descending the replica tree: the frozen field probability distribution
functions}
\label{app:FF}
In Section \ref{sec:SA} we have shown how spin averages can be computed using
a {\sl bottom-up} approach, that is starting from level $R$ at the bottom of 
the tree and climbing up towards level $0$ at the top of the tree.
A {\sl top-down} approach is also possible. 

To illustrate the procedure suppose we have to compute the following average,
\begin{equation}
\label{eq:d-ave}
  \langle g\rangle = \frac{
    \int {\cal D}_r(\alpha)\prod_{0}^{r} Z_r(\alpha)\ g_r(b^r_a)
                          }
                          {
    \int {\cal D}_r(\alpha)\prod_{0}^{r} Z_r(\alpha)
                          }
\end{equation}
where 
$g_r(b^r_a)$ is a generic function of the frozen field $b^r_a$ at scale $r$.
For example $g_r(b^r_a) = m_r(b^r_a)^2$ for the two-spin correlation, 
see eq. (\ref{eq:sa-m2i}).
The average (\ref{eq:d-ave}) can be rewritten in the simple form
\begin{equation}
\label{eq:d-avep}
  \langle g\rangle = \int dy\, P_r(y)\, g_r(y).
\end{equation}
by introducing the frozen field probability distribution function 
at scale $r$
\begin{equation}
\label{eq:d-Pr}
  P_r(y) = \frac{
    \int {\cal D}_r(\alpha)\prod_{0}^{r} Z_r(\alpha) \ 
    \delta(b^r_a - y)
                          }
                          {
    \int {\cal D}_r(\alpha)\prod_{0}^{r} Z_r(\alpha)
                          }
         =\langle \delta(b^r_a - y) \rangle .
\end{equation}
Following the procedure outlined in Section \ref{sec:SA} we can integrate 
the Gaussian variables $z^r_\alpha$ in eq. (\ref{eq:d-Pr}), ending with
\begin{equation}
\label{eq:d-Pr1}
  P_r(y) = \frac{
    \int {\cal D}_{r-1}(\alpha)\prod_{0}^{r-1} Z_{r-1}(\alpha) 
        \ g_{r-1}(b^{r-1}_a;y)
                          }
                          {
    \int {\cal D}_{r-1}(\alpha)\prod_{0}^{r-1} Z_{r-1}(\alpha)
                          }
\end{equation}
where 
\begin{equation}
\label{eq:d-gr1}
\fl
  g_{r-1}(b;y) = \frac{1}{\sqrt{2\pi\Delta\lambda_r}}
    \exp\left\{ -\frac{(y-b)^2}{2\Delta\lambda_r}
          + p_r\,\bigl[G_r(y) - G_{r-1}(b)\bigr]
        \right\}.
\end{equation}
Equation (\ref{eq:d-Pr1}) has the same structure as eq. (\ref{eq:d-ave}), thus
we can write
\begin{equation}
\label{eq:d-Pr1p}
 P_r(y) = \int d y' P_{r-1}(y') \  g_{r-1}(y';y)
\end{equation}
where $P_{r-1}(y)$ is given by eq. (\ref{eq:d-Pr}) with $r\to r-1$.
Inserting the expression (\ref{eq:d-gr1}) into (\ref{eq:d-Pr1p}) 
leads to the recursion relation
\begin{eqnarray}
\label{eq:d-Prrec}
  P_r(y) &=& \int Dz 
    \exp\left\{p_r\,\bigl[G_r(y) - G_{r-1}(\sqrt{\Delta\lambda_r}\,z + y)\bigr]
        \right\}
\nonumber\\
&\phantom{=}&
  \phantom{================}
    \times P_{r-1}(\sqrt{\Delta\lambda_r}\,z + y)
\end{eqnarray}
that gives the frozen field distribution at scale $r$ one it is known at scale
$r-1$. The initial condition is specified at level $0$, 
at the top of the tree, and reads
\begin{equation}
 P_0(y) = \frac{1}{\sqrt{2\pi\lambda_0}}
    \exp\left\{ -\frac{(y-b)^2}{2\lambda_0}
        \right\}.
\end{equation}
as follows from eqs. (\ref{eq:d-Pr1}) and (\ref{eq:d-gr1}).

In the limit $R\to\infty$ $P_r(y) \to P(x,y)$ and
the recursion relation (\ref{eq:d-Prrec}) is replaced by the partial 
differential equation
\begin{equation}
\label{eq:d-Pxy}
  \dot{P}(x,y) = \frac{\dot{\lambda}(x)}{2}
  \Bigl[ P''(x,y) - 2\,x\,\bigl[m(x,y)\,P(x,y)\bigr]'
  \Bigr]
\end{equation}
where $m(x,y) = G'(x,y)$, with the initial condition at $x=0$:
\begin{equation}
\label{eq:d-P0y}
 P(0,y) = \frac{1}{\sqrt{2\pi\lambda(0)}}
    \exp\left\{ -\frac{(y-b)^2}{2\lambda(0)}
        \right\}.
\end{equation}
Equations (\ref{eq:d-Pxy}) and (\ref{eq:d-P0y}) where first derived
by Sommers and Dupont \cite{SomDup84} using a variational approach.
We note that taking $g_r(b) = m_r(b)^2$ from eq. (\ref{eq:d-avep}) 
one recovers in  limit $R\to\infty$ the Sommers-Dupont expression
\begin{equation}
 Q(x) = \int dy\, P(x,y)\, m(x,y)^2.
\end{equation}

The approach in terms of frozen field distribution functions can be 
generalized to deal with averages of quantities that depend on more then 
one local field.
Suppose for example that $g_r\to g_r(b^r_a, b^r_b)$ with $a\not=b$.
In this case eq. (\ref{eq:d-avep}) is replaced by
\begin{equation}
 \langle g\rangle = \int dy_1\,dy_2\, P_r(y_1,y_2)\, g_r(y_1,y_2)
\end{equation}
where $P_r(y_1,y_2)$ is the probability distribution function of the 
frozen fields $y_1$ and $y_2$ lying on two different branches 
of the tree at scale $r$. 
This satisfies the {\sl top-down} recursion relation
\begin{eqnarray}
\fl
\label{eq:d-Prrec2}
  P_r(y_1,y_2) &=& \int Dz_1\, Dz_2
  \prod_{k=1,2}
    \exp\left\{p_r\,
         \bigl[G_r(y_k) - G_{r-1}(\sqrt{\Delta\lambda_r}\,z_k + y_k)\bigr]
        \right\}
\nonumber\\
\fl
&\phantom{=}&
  \phantom{============}
 \times P_{r-1}(\sqrt{\Delta\lambda_r}\,z_1 + y_1, 
                \sqrt{\Delta\lambda_r}\,z_2 + y_2).
\end{eqnarray}
The initial condition is specified at the branching point $s < r$ 
where the two branches meet, and reads
\begin{equation}
    P_r(y_1,y_2)\Bigr|_{r=s} = \delta(y_1 - y_2)\, P_s(y_1).
\end{equation}
It is easy to verify that the frozen field distribution functions 
obey the sum-rule 
\begin{equation}
 \int dy_2\, P_r(y_1, y_2) = P_r(y_1), \qquad \forall r.
\end{equation}
In the continuous limit the recursion relation (\ref{eq:d-Prrec2})
is replaced by the partial differential equation
\begin{equation}
\fl
\label{eq:d-Pxyy}
  \dot{P}(x,y_1,y_2) = \frac{\dot{\lambda}(x)}{2}
  \sum_{k=1,2}
  \Bigl\{ \partial_k^2 P(x,y_1,y_2) 
         - 2\,x\,\partial_k\bigl[m(x,y_k)\,\partial_kP(x,y_1,y_2)\bigr]
  \Bigr\}
\end{equation}
where $\partial_k = (\partial/\partial y_k)$. The generalization to frozen
field distribution functions of any number of independent frozen fields is
straightforward.

\end{document}